\documentclass{emulateapj}
\usepackage{graphicx}
\usepackage{url}

\shorttitle{Flare continuum opacity}
\shortauthors{Potts et al.}

\begin{document}

\title{The optical depth of white-light flare continuum}

\author{Hugh Potts, Hugh Hudson\altaffilmark{1}, Lyndsay Fletcher, and Declan Diver}
\affil{Department of Physics and Astronomy, University of Glasgow, G12 8QQ, UK}
\email{hugh@astro.gla.ac.uk}
\altaffiltext{1}{also SSL, UC Berkeley, CA, USA 94720}


\begin{abstract}
The white-light continuum emission of a solar flare remains a puzzle as regards its height of formation and its emission mechanism(s).
This continuum, and its extension into the near~UV, contain the bulk of the energy radiated by a flare, and so its explanation is a high priority.
We describe a method to determine the optical depth of the emitting layer and apply it to the well-studied flare of 2002 July~15, making use of MDI pseudo-continuum intensity images.
We find the optical depth of the visible continuum in all flare images, including an impulsive ribbon structure to be small, consistent with the observation of Balmer and Paschen edges in other events.
\end{abstract}

\keywords{Sun: flares --- Sun: photosphere}

\section{Introduction}\label{sec:intro}
The first-observed feature of a solar flare, the ``white light'' continuum, remains enigmatic to the present day.
Nevertheless it (and the related near-UV continuum) contain the bulk of a flare's radiant energy,
so that an understanding of how it forms would help a great deal in our understanding of flare 
physics.
Indeed, understanding the most important component of the energy of a flare must in fact be the single most important problem in flare physics \citep[e.g.,][]{1989SoPh..121..261N}.

The difficulties limiting our observational knowledge are several.
First, the continuum excess over the bright photospheric emission only rises to some tens of percent even in the most powerful flares, and there are both practical (the image contrasts of sunspots) and intrinsic (convection and other solar ``noise,'' such as the p-modes) limits to the photometric precision.
Also the strong variability both in space and in time of the white-light emission has made it difficult to obtain a good characterization of  the spectrum of the emission.  
The data that do exist suggest two types of white-light flare continuum: those associated with the impulsive phase and which correlate with particle signatures, and more gradual brightenings which have featureless continuum (Boyer et al., 1985)\footnote{These classes are also called type~I and type~II (Machado et al., 1986).}.\nocite{1986A&A...159...33M}
\nocite{1985SoPh...98..255B}

We do know that the white-light continuum can have a strong association with the hard X-ray emission in the impulsive phase of the flare, both temporal \citep{1975SoPh...40..141R,1992PASJ...44L..77H,1993SoPh..143..201N} and spatial \cite[e.g.,][]{2003ApJ...595..483M,2006ApJ...641.1210X}.
This indeed was one of the first motivations for the ``thick target model'', which associates the chromospheric and photospheric effects of a flare with the energy losses of high-energy particles \citep{1970SoPh...15..176N,1970SoPh...13..471S}, and specifically the electrons in the 10-100~keV range \citep{1972SoPh...24..414H}.
The energetics of the electrons matches well \citep{1972SoPh...24..414H,2007ApJ...656.1187F}, at least to the extent that we understand the energy in the white-light and UV continuum.
Recent observations of the bolometric luminosity of a solar flare \citep{2004GeoRL..3110802W,2008cosp...37.1617K,2010arXiv1003.4194Q} have proven to be consistent with the idea that the white-light continuum and its UV~extension dominate the flare luminosity, and furthermore that this luminosity appears in the impulsive phase.
Thus mechanisms that can explain white-light continuum formation must have a close link to the non-thermal electrons responsible for hard X-ray emission in the impulsive phase.

The theoretical quandary that has blocked a full understanding is that the 10-100~keV electrons of the impulsive phase cannot penetrate as deeply as the non-flaring visible photosphere (the level for which the optical depth at 5000\AA~$\tau_{5000}$ = 1, where one might naively expect the white-light flare continuum to originate).
This problem was exacerbated in the extreme when \cite{2004ApJ...607L.131X} found significant contrast for flare emission even at 1.56$\mu$, nominally the ``opacity minimum'' region of the spectrum, which according to standard modeling would form actually \textit{below} 
$\tau_{5000}$~=~1.
Several possible solutions to this quandary have been proposed, viz:
\begin{enumerate}
\item Over-ionization in the chromosphere: at the stopping depth of electrons of sufficient total energy, excess ionization can enhance the continuum adequately across the spectrum \citep{1972SoPh...24..414H,1986A&A...156...73A};
\item Radiative back-warming:  hydrogen recombination radiation frpm the primary stopping height of the thick-target electrons heats the photosphere itself sufficiently to produce the observed continuum \citep{1989SoPh..124..303M};
\item Proton energy losses:  protons at a few MeV energy penetrate more deeply \citep{1970SoPh...15..176N};
\item Wrong model: the thick-target model does not apply, and both the hard X-ray emission and the visible continuum come from a deeper layer excited by some other mechanism \cite[e.g.][]{1968QJRAS...9..294U}.
\end{enumerate}

Of these possibilities items (1) and (2) make clear predictions for the optical depth of the visible continuum; an overionized layer in the chromosphere will produce an optically thin spectrum of recombination radiation, while backwarming would more closely have a blackbody spectral distribution.
What spectroscopy does exists favors (1) to a certain extent, because the best impulsive-phase
optical spectroscopy suggests the presence of the Balmer jump and even the Paschen jump \citep{1984SoPh...92..217N}.
The other categories of explanation (3) and (4) are more problematic since so little theoretical work has been done, but it is clear that the classical thick-target model needs revision because the modern data require beam intensities greater than seem physically plausible \citep{2008ApJ...675.1645F,2009A&A...508..993B}.

The backwarming mechanism deserves special mention, because as \cite{1989SoPh..124..303M} point out, it \textit{must} happen to a certain extent: Balmer radiation from the chromosphere emitted downwards can penetrate freely to the upper photosphere (the temperature-minimum region),  heating the medium up to radiative equilibrium.
Then the usual H$^-$ opacity at a slightly elevated temperature (say 300~K) could produce the white-light flare.
This process would be independent of the optical depth of the primary emitter in the chromosphere and, if it were small, would permit the Balmer and Paschen edges to appear in the spectrum (though diluted).
The logic behind this theory rests upon steady-state 1D~atmospheric modeling, so that if 2D~spatial structure on the scale of the chromosphere/photosphere height difference exists, spatial structure will appear in the backwarmed photosphere; similarly time variations would introduce discrepancies between the chromospheric UV source and the photospheric response.
In a backwarmed photosphere there would be no spatial structure on scales smaller than that of the photosphere/chromosphere height separation.

In this paper we address the essential property of optical depth directly, by analysis of SOHO/MDI\footnote{The Michelson-Doppler Imager \citep{1995SoPh..162..129S} on board the Solar-Heliophysical Observatory spacecraft.} images of the well-observed flare of 2002 July~15.
The essence of the technique is simply to correlate the known intensity structure of the photosphere at the location of the flare brightening with the brightening itself, and to interpret this in terms of a simple slab model for the radiative transfer (Boyer et al. 1985).
To our knowledge no analysis of this kind has been carried out before.

\section{The X3 flare of 2002~July~15}

\begin{figure}
\epsscale{1}
\plotone{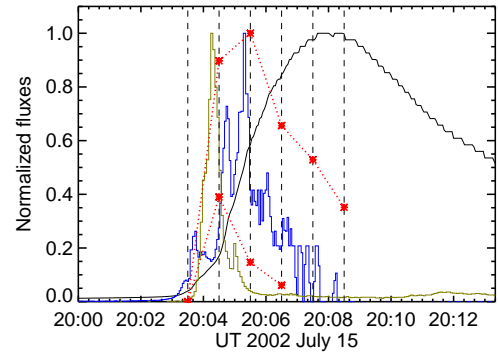}
\caption{Time histories for the flare of 2002 July~15, normalized to the individual maxima.
The thin black line is GOES 1-8\AA, and the blue line its time derivative. 
The red points are white-light fluxes, total (upper) and the ribbon-like feature only (lower), both scaled to the peak of the total.
The orange line is the OVSA 18~GHz flux density.
}
\label{fig:ts}
\end{figure}

This flare (GOES class X3) famously showed a multi-turn helical structure formed in its plasma ejection, as observed in TRACE~UV images thought to represent mainly C{\sc iv} emission \citep{2003ApJ...593L.137L}.
The chromospheric and photospheric properties of the event were recorded with Mees Observatory  imaging spectroscopy in H$\alpha$, and with white-light imagery from the Imaging Vector Magnetograph at Mees \citep{2005ApJ...620.1092L}.
Unfortunately RHESSI was at orbit night during the impulsive phase, so there are no hard X-ray images available, but Owens Valley microwave data were available.
Figure~\ref{fig:ts} arrays the flare white light, soft X-ray, and OVSA 1-18~GHz microwave emissions as time series for reference; Li et al. have estimated energy fluxes for this comparison.
To define the impulsive phase Figure~\ref{fig:ts} also shows the GOES time derivative, since RHESSI hard X-ray data were not available.

\begin{figure*}
\plotone{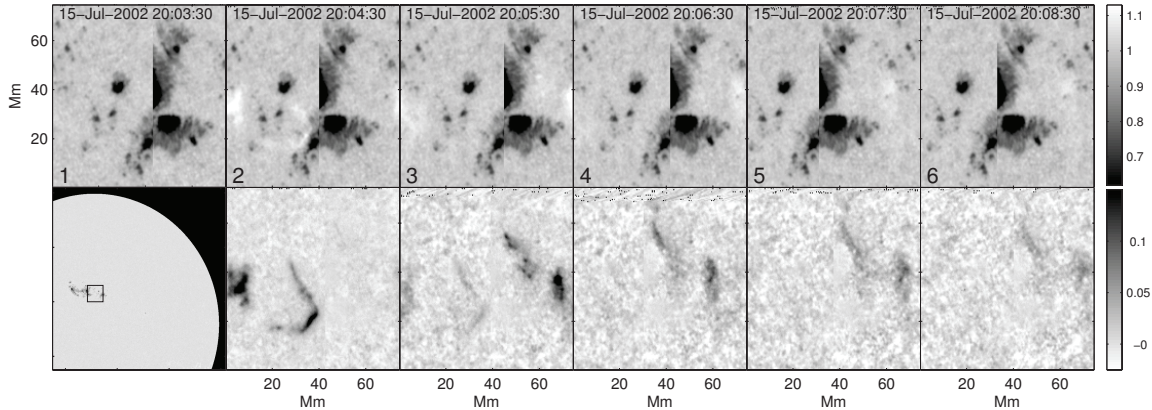}
\caption{The time evolution of the flare at one-minute cadence. 
The upper images show the flatfield-corrected MDI continuum images, and the lower images show the difference between these and a background photospheric image reconstructed by interpolation, color table reversed for clarity.}
\label{fig:Frames}
\end{figure*}

The data we utilize in this study are the MDI ``pseudo-continuum'' intensity data, actually narrow-band samples of the continuum near the magnetically sensitive photospheric absorption line of Ni{\sc i} at 6768\AA~(Ding et al. 2003).\nocite{2003A&A...403.1151D}
The basic data are 1024$\times$1024 time-averaged images at 1-minute cadence in a 10.5$'$ square field of view at disk center (pixel size 0.615$''$; diffraction limit 1.22$\lambda$/D = 1.36$''$).
The telescope has excellent pointing stability and the noise in a given pixel is predominantly solar in origin -- broad-band variations from convective motions, plus the p-modes. 
The photometric accuracy of the data was further improved by applying the secondary flatfields detailed in the MDI flatfield repository \citep{2009SoPh..258..343P}
\footnote{\url{http://soi.stanford.edu/sssc/MDI\_continuum\_hr\_flatfields/ flatfields.html}.}.
Figure~\ref{fig:Frames} shows the flatfield-corrected images.

Our analysis focuses initially on the crescent-shaped flare region visible clearly in the 20:30~UT image (Frame~2 of Figure~\ref{fig:Frames}). 
We then extend the technique to a generalised case. 
This part of the 15~July flare is particulary easy to analyse as it has the advantage of passing over regions with a wide range of photospheric intensities, and it proves simple to construct a low-noise photospheric background for it.
The continuum emission has an elongated structure which might be misinterpreted as a loop in projection, but which the multiwavelength observations of Li et al. (2005) show clearly to be the eastern ribbon of the flare (see their Figure~7, in particular the frames showing the H$\alpha$ blue wing at 20:03:32 and 20:04:01~UT for reference). 
The Li~et~al. white-light data show a good match to the initial bright microwave source (the orange line in Figure~\ref{fig:ts}), which our less-frequent sampling misses.

Distributions of the intensity along an arc defining the midpoint of this ribbon source, and along the same path on the preflare image (one minute prior) are shown in Figure~\ref{fig:FlareXsect}; these will be used together with a simple model to argue that the flare optical depth is small.
However a first indication that this is the case is simply that the photospheric structure is clearly superposed on the flare image.

\section{Simple model}\label{sec:simple}

Our simple model of the radiative transfer in the white-light flare consists of a homogeneous slab (optical thickness $\tau$, source function S$_1$) located above an optically-thick photosphere at temperature T$_0$.
This analysis follows that of Boyer et al. (1985)\nocite{1985SoPh...98..255B}.
We assume that the photosphere (in a given pixel) has intensity $I_0=S_0$ before the flare, where $S_0$ is the source function (the Planck function if in LTE).
Here we explicitly ignore backwarming (see Section~{\ref{sec:intro}), assuming that the photosphere does not change significantly during the flare.
As will be seen, this does not contradict the data we discuss for this flare.
The observed intensity of a given pixel during the flare is $I_F$, which in this model consists of  a combination of (attenuated) photospheric emission and direct flare emission. 
The observed brightness during the flare is given generally by
\begin{equation}
I_F=S_0 e^{-\tau}+S_1(1-e^{-\tau})
\label{eq:FlareInt}
\end{equation}
where $S_1$ is the source function for the flaring emission layer.

We illustrate the analysis for  $\tau$ with a simulated white-light flare resembling the part of
the event we analyze below. 
The model photosphere contains weak modulations patterned after granules, and a dark `pore' region.
Consider the intensity profiles of the model flare brightening as shown in Figure~\ref{fig:DemoFlare}. 
This shows the spatial cross section of three simple examples, each with the same Gaussian flare emission profile (upper right panel in red), but with different optical depths in the slab and thus different emergent intensities (green). 
Each `flare' is arranged so that the flare brightening has a 20\% increase relative to the mean photosphere. 
The lower graphs show the total emission before (blue) and during (green) the flare, and the upper graphs (red) show the increase in intensity caused by the flare. 
In the case with large optical depth ($\tau=1$) the photospheric emission is significantly attenuated when observed through the flare, and so appears with a lower amplitude superimposed on the flare emission. 
If the photospheric emission is subtracted from the total flare emission then the result is a combination of the flare emission with an inverted image of the photosphere, having a negative correlation with the original photosphere. 
The right-hand column shows the case where the flare has very small optical depth ($\tau=0.01$). 
In this case the total emission during the flare is approximately the sum of the unattenuated  photospheric emission and the flare emission. 
When the photospheric background is subtracted from the flare, the result is the true flare intensity profile, with no inverted component from the photosphere. 
It can be seen from this that if it is possible to know the intensity of the photosphere underlying the flare then the optical depth of the flare can be determined directly from the degree of spatial correlation between the flare brightening and the background image.

\begin{figure*}[h]
\epsscale{.8}
\plotone{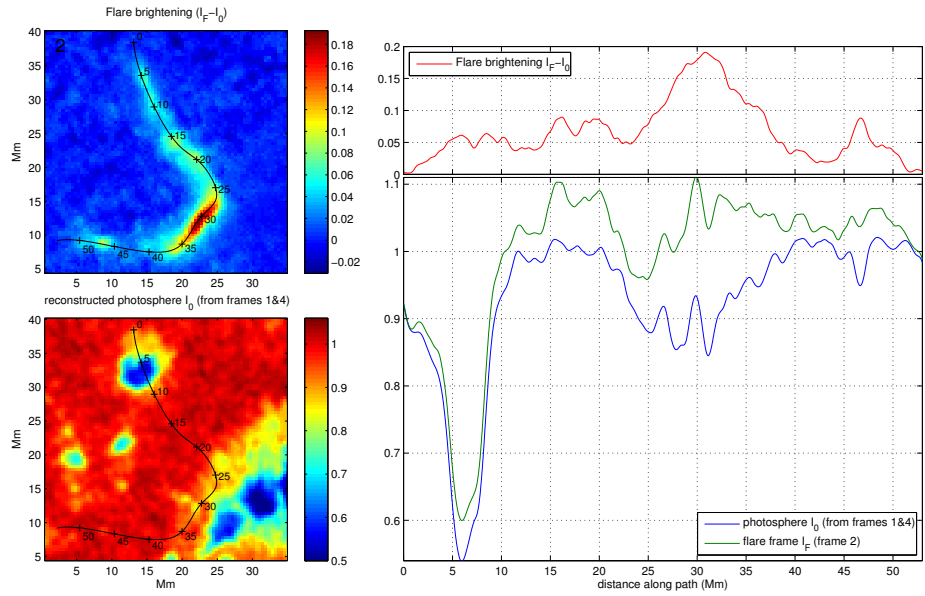}
\caption{Intensity profile along the selected part of the flare, based on Image Frame 2 (20:04:30~UT).
\textit{Left upper,} the flare brightening excess; \textit{left lower,} the photospheric structure underlying the emission; \textit{right,} traces along the structure.
Here green shows the flare brightening, blue the reconstructed photosphere, and red the flare excess.
}
\label{fig:FlareXsect}    
\end{figure*}

This is reinforced by the upper graph, where the photospheric intensity is subtracted from the flare frame; this is  also shown as an image in the lower left panel. 
If the flare were significantly optically thick then the variations on the attenuated photosphere viewed through the flare would be smaller than those viewed directly, and therefor this difference image would contain an inverted represention of the photosphere, as can be seen in the middle column of Figure~\ref{fig:DemoFlare}. The absence of this tells us that the optical depth $\tau$ must be small, and furthermore that the background photosphere has not varied drastically during this minute.

The analysis aims at determining the quantities $\tau$ and S$_1$~=~bB(T$_1$) at each pixel of the continuum brightening; b~is the departure coefficient and B(T$_1$) the Planck function at the source layer temperature T$_1$.
In general, assuming LTE \citep[local thermodynamic equilibrium; see e.g.][]{1978stat.book.....M}, the opacity and source function are related by Kirchoff's law I~=~$\kappa$S.
If LTE held and b~=~1 we could deduce the two unknowns for each pixel directly.
In practice we cannot assume LTE, unfortunately; indeed the phenomenon of white-light continuum emission in a flare requires a highly atypical application of radiative-transfer theory to the physics of the solar atmosphere
because of the extreme conditions.
The relaxation of the LTE condition for the continuum would mean that an additional unknown equivalent to the non-LTE departure coefficient must be introduced.
We deal with this here by using the heuristic approximation described in the following section.

\begin{figure}
\epsscale{1.00}
\plotone{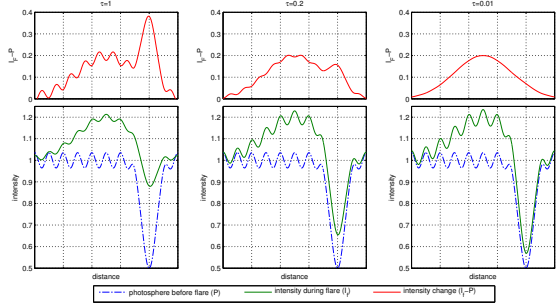}
\caption{Simulated white-light flare emission profiles compared to the underlying photospheric emission for different values of the optical depth ($\tau$). 
The variations on the photosphere represent granules and a small sunspot, and all intensities are normalised to the mean photospheric intensity.
Color coding is the same as in Figure~\ref{fig:FlareXsect}.
}
\label{fig:DemoFlare} 
\end{figure}

\section{Analysis}

\subsection{Heuristic opacity model}

We note that an idealized backwarming model for the flare brightening would also match this observation, but we can rule this out because of the fine structure present in the flare image.
\cite{2007PASJ...59S.807I} discuss this aspect of white-light flare geometry in detail.
In our model, the height of the slab is comparable to the width of the ribbon-like flare emission, 
which varies substantially along its length and is unresolved in its narrowest parts.
Furthermore, the flare region we discuss has a lifetime less than 60~s, whereas the cooling time 
in the VAL\_C model \citep{1981ApJS...45..635V} is about 80~s at the photosphere;
we calculate the timescale as $\varepsilon$H/${\mathcal L}_\odot$, where $\varepsilon$, H, and ${\mathcal L}_\odot$ are the thermal energy density, scale height of the photosphere, and photospheric luminosity.
This is not a decisive argument because of the low time resolution of our observations.

In the case that $\tau$ is small, we can simplify equation \ref{eq:FlareInt} by taking a linear expansion of the exponential function $e^{-\tau}\approx 1-\tau$ :
\begin{equation}
I_F \approx S_0 (1-\tau)+S_1\tau
\label{eq:int_linear}
\end{equation}
This approximation gives errors of less than 4\% for $\tau < 0.25$, and less than 0.5\% for $\tau<0.1$.

\begin{figure}
\plotone{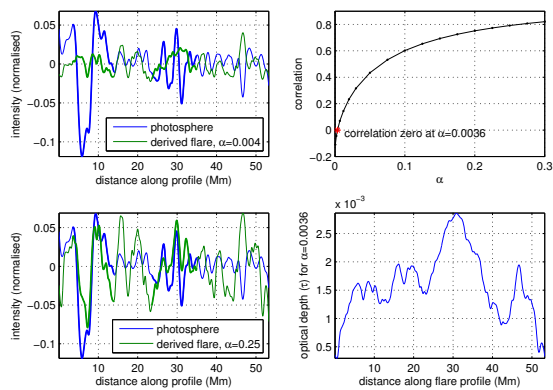}
\caption{ \textit{Left:} Calculated flare intensities for $\alpha=0.004$ and $\alpha=0.25$ for the path shown in Figure~\ref{fig:FlareXsect}, after passing through a high-pass filter with cutoff of 3.5~Mm FWHM.
The bold line sections are the regions selected for the correlation calculations where significant flare brightening is combined with large photospheric variation. Notice the correlation between the derived flare intensity and the photosphere intensity for the $\alpha=0.25$ case.
\textit{Right:} Correlation of the derived flare intensity with the underlying photospheric emission as a function of the opacity parameter~$\alpha$. The correlation is zero for $\alpha=0.0037$, and the lower panel shows the optical depth of the derived flare  for this value of $\alpha$.}
\label{fig:Correlations} 
\end{figure}

As both $\tau$ and $S_1$ are unknown Equation~\ref{eq:int_linear} cannot be solved from just the knowledge of $I_F$ and $S_0$. 
We therefore need to postulate an heuristic relationship between $\tau$ and $S_1$, which would be related to the physics of the emission and absorption in the slab region to make the problem tractable.
If the physical situation in the source region (the slab) is characterized by overionization \citep{1972SoPh...24..414H} and heating, we would expect $\tau$ and $S_1$ to correlate with each other, although the overionization would tend to increase the departure coefficient for the continuum, and hold the temperature T$_1$ to a lower level.
The simplest relationship would be $\tau$~=~const., but this is clearly too restrictive; it would require that faint flares perfectly match the photosphere.

We have therefore adopted $\tau=\alpha(S_1-S_0)$ as the simplest functional dependence of 
opacity on the physical conditions in the source.
This is consistent with our physical argument regarding the effects of non-thermal particles heating the flare region. 
Particle heating would generally be expected to increase both the temperature and the ionisation faction, leading to strong continuum radiation, e.g. in the Balmer free-bound continuum. 
The total energy radiated from the flare volume when the optical depth is small would then be proportional to the physical depth of the heated region and the source function in that region. 
The net radiation which is observed will fall between the difference between this and the underlying photospheric radiation intensity, hence the relation above.

With this heuristic relationship, we find a quadratic equation for S$_1$ and solve it as
\begin{equation}
S_1 = S_0 + \sqrt{{{I-S_0}\over{\alpha}}}.\label{eq:s1}
\end{equation}
The value of the constant~$\alpha$, which sets the optical depth, is unknown, but we can determine it by considering the spatial distribution of the flare emission.
Note that~$\alpha$ is not the optical depth itself, but a parameter describing its physics.
As can been seen from Figure~\ref{fig:FlareXsect}, the photosphere has more structures at small scale than the flare brightening. 
Recall also our assertion that the flare intensity is not correlated with the underlying photospheric intensity before the flare.
The flare source function $S_1$ can be calculated using Equation~\ref{eq:FlareInt} for a range of values of 
$\alpha$ and hence $\tau$. 
If the variation of this derived flare source function at small scales is compared to the intensity of the underlying photosphere interpolated between preceding and following image frames, the two should be uncorrelated (except for random coincidence) when the correct optical depth is used in the calculation.

\subsection{Estimation of the opacity parameter $\alpha$}
\begin{figure*}
\plotone{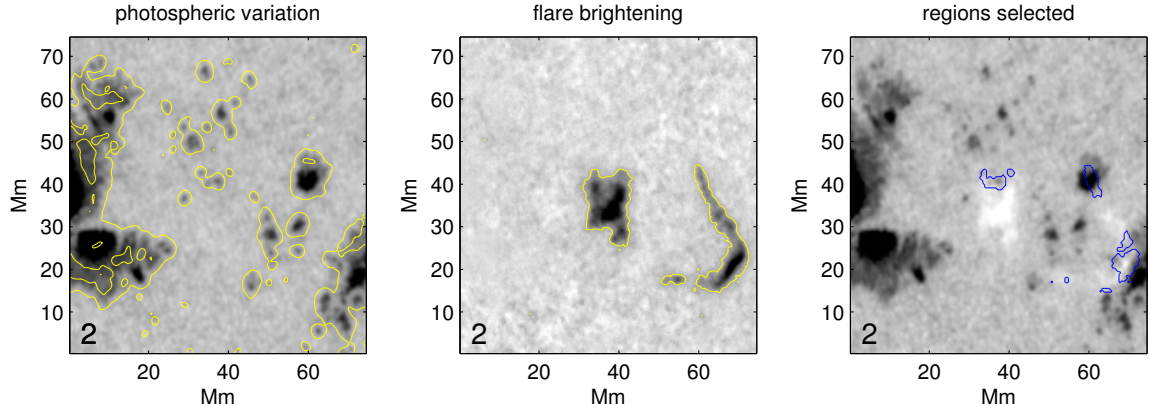}
\caption{Selection of regions for analysis. 
\textit{Left:} Map of the photospheric RMS variability; \textit{middle:} difference image showing the flare brightening, with contours showing the selected regions;  \textit{right:} the union of these selections for analysis.
The label ``2'' identifies Image Frame~2 at 20:04:30~UT.
}
\label{fig:Selections} 
\end{figure*}

The key to the method described above it to observe the attenuation of the photospheric variation when viewed through the flare. 
In order to do this we need to identify regions where the following are true:

\begin{enumerate}
    \item We must be able to determine the underlying photospheric emission at the time of the flare. 
    This requires reconstructing the photosphere by interpolating surrounding non-flare image frames.  
    The error on this can be estimated by comparing the reconstruction to the observed photosphere in regions outside the flare.
    \item There must be significant spatial variation in the photospheric intensity over the flare region, which is much larger than the error on our photospheric reconstruction.
    \item There must be significant flare brightening, again, much larger than the noise on the photospheric reconstruction.
   \item The variation of the photospheric intensity needs to be large as possible in comparison to the co-spatial intensity variation of the flare at the scales considered. 
This is necessary to reduce the effects of chance correlations
\end{enumerate}

The flare cross section shown in Figure~\ref{fig:FlareXsect} is a particularly good example that
meets these criteria. 
In this case the reconstructed photospheric background was particularly accurate as it could be generated by a linear interpolation between image frames 3 and 6 (see figure \ref{fig:Frames}) due to the rapid evolution of the flare brightening. 
The error on this reconstruction was  $\sim$0.5\% of the mean photospheric intensity, estimated from the scatter of the reconstructed photosphere in regions away from the flare to observations. 
The profiles through this ribbon are shown in the right-hand panel of Figure~\ref{fig:FlareXsect}. 
The photospheric variations are generally at a smaller scale than the flare brightening, so to exploit this we passed the data though a high-pass filter with cut-off of 3.5~Mm FWHM, and chose the areas of high photospheric variation manually.
These are shown by the bold portions of the filtered cross section shown in 
Figure~\ref{fig:Correlations}. 
A flare model profile was then generated using the reconstructed S$_0$ and Equation~\ref{eq:s1} for a range of different values of the opacity parameter $\alpha$, and the correlation between this and the photospheric background was calculated. 
Figure~\ref{fig:Correlations} shows the results, with reconstructions for $\alpha=0.25$ and $\alpha=0.004$ shown in the left panels, and the variation of the correlation with $\alpha$~shown on the right hand panel.
For the case $\alpha=0.25$ the generated flare profile shows a significant positive correlation with the underlying photosphere, which completely disappears when when $\alpha$ is small.
According to the arguments in Section~\ref{sec:simple}, this implies small optical depths. 
In this case the correlation goes to zero when $\alpha=0.0036$. 
From our assumption that $\tau=\alpha(S_1-S_0)$ this gives an average optical depth for the flare of $\tau=0.0016$, with the variation of $\tau$ along the path shown in the the lower panel.
This results from the low photospheric noise.

\begin{table}
\caption{Summary of the results from different image frames
   }
\begin{tabular}{ c | c | c | c | c }
  Frame & $\alpha_0$ & Mean $\tau$   & Photospheric & $\Delta$S/S (\%)$^a$  \\
  &  &    & rms error &   \\
\hline
2 & 0.0037 &    0.015 &    0.78  &     8.5   \\
3 & 0.0131 &    0.033 &    1.17  &    11.5   \\
4 & 0.0309 &    0.042 &    1.28  &     7.2   \\
5 & 0.0119 &    0.024 &    1.31  &     5.9   \\
6 & 0.0233 &    0.032 &    1.35  &     5.2   \\
7 & 0.0066 &    0.016 &    1.32  &     4.7   \\
all frames & & 0.028 & & \\
\hline
  \end{tabular}
  
$^a\Delta$S/S is the flare excess brightness, from the selected regions, normalized to the non-flaring photosphere.
\label{tab:results}
\end{table}

The above example indicates very small optical depth, but since the method assumes no correlation between the detailed photospheric intensity and flare brightening, random correlations could also give that result. 
To test this we need to look at a larger data set. 
To automate this algorithms were constructed that isolated regions where the four criteria listed above were true. 
An example of this from Frame~2 is shown in Figure~\ref{fig:Selections}. 
This process was carried out for all image frames and the results are summarised in 
Table~\ref{tab:results}. 
As can be seen, the mean optical depth of the flares in each frame is small, with a maximum of 0.042. Notice that the value of $\tau$ is not related to the average flare brightening, and is lowest for the frame which has the lowest photospheric error (Image Frame~2).

\subsection{Error analysis}

\begin{figure}
\plotone{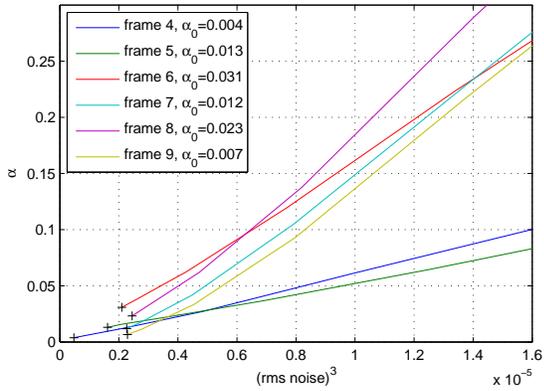}
\caption{The effect of noise in the photospheric reconstruction on the derived value for $\alpha$. Extrapolating this graph back to the \emph{y} axis allows the zero noise case to be estimated. }
\label{fig:Noise}
\end{figure}

The major error source in the above procedure is the error in the estimate of the photospheric brightness in regions under flare areas.
Although the average of this error will be zero by assumption, the spurious structure it introduces to the photospheric reconstruction gives a systematic error to the derived value of $\alpha$, and hence to the derived optical depth, always tending to increase the derived value. 
To see why this should be true consider the effects of adding a pattern of random noise to the reconstruction of the non-flare photosphere. 
The observed intensity of the solar surface with the flare present will not contain this error term, so it will appear that the flare is masking it, suggesting that it is more optically thick.

In order to evaluate the significance of the photospheric noise, we add Gaussian random noise of various (known) amplitude levels to the reconstructed photosphere. 
First the RMS error on the photospheric reconstructions was measured for each image frame by comparing the reconstructed photospheric image in a region away from the flare emission to the observed photosphere; results are shown in Table~\ref{tab:results}. 
Normally distributed noise with a range of known standard deviations was then added to the reconstructed photosphere and the calculation of~$\alpha$ repeated. 
This was done multiple times with different random noise for each frame and each value of the noise amplitude, in order to avoid random correlations between added noise and the photosphere. 
The variation in the derived value of~$\alpha$ versus the total photospheric error for each frame is shown in Figure~\ref{fig:Noise}. 
It can be seen that as the error in the photospheric noise increases, the derived value of~$\alpha$ also increases as expected. 
Note that the values of $\alpha$~determined in image frames where the flare brightness is larger (see Table~\ref{tab:results}) are much less sensitive to the effect of the noise.

Using this graph it is possible to extrapolate back to a `zero error' value for~$\alpha$. 
If this is done for each frame using a simple polynomial extrapolation we get an average value for all the data of $\alpha=-0.0001\pm0.01$, so to within the accuracy of these measurements the optical depth is effectively zero.

\subsection{Discussion}

Our direct analysis of the flare emission from seven image frames, comprising a total of 1200 pixels where the flare emission significantly enhanced the surface brightness, has given us the result that the optical depth is $0.028\pm0.01$. 
This value however should be regarded as an upper limit on the opacity of the flaring regions; when the effect of photospheric noise is considered the optical depth becomes too small to measure, and certainly less than 0.01.

As a result of this the assumption of our heuristic opacity model that~$\tau$ is depends on the flare source function becomes unimportant, as the optical depth is so close to zero that a more appropriate model for the emission becomes $
I_F=S_0+I_1,$
where I$_1$ is the flare emission.
The flare excess simply adds to the photospheric emission.

\section{Conclusions}

The optical depth of the white-light flare regions we have studied is very small, indistinguishable from zero in this study, and in any case less than $\sim$0.01.
We infer from this that the flare must be of low density and hot, almost certainly far from LTE.
The temperature cannot be determined from this but is generally constrained by the new flare bolometric observations \citep{2004GeoRL..3110802W,2008cosp...37.1617K,2010arXiv1003.4194Q}; many authors suggest a value near 10$^4$~K (e.g., Hudson et al. 2010).
\nocite{2010arXiv1001.1005H}
These results definitely tend to reduce the importance of photospheric backwarming in our understanding of white-light flare emission.
First, the low density and high temperature imply an emission source high in the atmosphere, consistent with stopping depths of low-energy electrons but inconsistent with stopping depths of the electrons required for backwarming models \citep{2007ASPC..368..423F}.
Second, the flare image scales (in this case limited by MDI resolution) are of order 1~Mm. 
If backwarming contributed significantly to the emission then the maximum height of the emission must be around half of the feature size (simple geometric ray model), which would be too low for an optically thin case \citep[see][]{2007PASJ...59S.807I}.

Because the MDI data represent averages over one-minute intervals, and because they represent only a narrow slice of the true continuum, this result should be considered as a preliminary one.
Newer data with better image cadence and spatial resolution (Hinode or SDO in space, or a variety of ground-based instruments leading up to ATST) should be applied to this interesting problem.
If confirmed, this result suggests that the generally accepted picture of flare energy storage in the corona, with flare effects in the lower atmosphere derived from this energy reservoir, must be correct.
Note that this is the usual assumption, but that it has not been easy to establish observationally.

\bigskip
{\bf Acknowledgments:} This work is supported by the EU's SOLAIRE Research and Training Network at the
University of Glasgow (MTRN-CT-2006-035484) and by Rolling Grant ST/F002637/1 from
the UKÕs Science and Technology Facilities Council. 
HH also thanks NASA for support under contract NAS 5-98033 for RHESSI.

\bibliographystyle{apj}
\bibliography{opacity}
\date{\today}

\end{document}